\documentclass[usenatbib,useAMS]{mn2e}

\usepackage{epsf}
\usepackage[dvips]{epsfig}
\usepackage{times}

\begin{document}

\label{firstpage}

\title[Merging sub-clumps]{Merging time-scales of stellar sub-clumps in
  young star-forming regions} 

\author[Fellhauer et al.]{
  M. Fellhauer$^{1,2}$, M. I. Wilkinson$^{3}$, P. Kroupa$^{4}$\\
$^{1}$ Institute of Astronomy, University of Cambridge, Madingley Road,
Cambridge CB3 0HA, UK \\
$^{2}$ Departamento de Astronomia, Universidad de Concepcion, Casila
160-C, Concepcion, Chile \\
$^{3}$ Dept.\ of Physics and Astronomy, University of Leicester,
University Road, Leicester LE1 7RH, UK \\
$^{4}$ Argelander-Institut f\"{u}r Astronomie, Universit\"{a}t Bonn, Auf
dem H\"ugel 71, 53121 Bonn, Germany
} 

\pagerange{\pageref{firstpage}--\pageref{lastpage}} \pubyear{2009}

\maketitle

\begin{abstract}
  Recent observations and hydrodynamical simulations of star formation
  inside a giant molecular cloud have revealed that, within a star
  forming region, stars do not form evenly distributed throughout this
  region, but rather in small sub-clumps.  It is generally believed
  that these sub-clumps merge and form a young star cluster.  The
  time-scale of this merging process is crucial for the evolution and
  the possible survival of the final star cluster.  The key issue is
  whether this merging process happens faster than the time needed to
  remove the residual gas of the cloud.  A merging time-scale shorter
  than the gas-removal time would enhance the survival chances of the
  resulting star cluster.  In this paper we show by means of numerical
  simulations that the time-scale of the merging is indeed very fast.
  Depending on the details of the initial sub-clump distribution, the
  merging may occur before the gas is expelled from the newly-formed
  cluster either via supernovae or the winds from massive stars.
  Our simulations further show that the resulting merger-objects have a 
  higher effective star formation efficiency than the overall star 
  forming region and confirm the results that mass-segregated 
  sub-clumps form mass-segregated merger-objects.
\end{abstract}


\section{Introduction}
\label{sec:intro}

The formation and survival of young, embedded star clusters is an
important issue not only for the star cluster community but
also has implications for our understanding of the formation and
evolution of galaxies as a whole.  Stars are now believed to form
mostly in star clusters, which later dissolve and distribute their
stars to their host galaxy.

In recent years, considerable progress in our understanding of the
star formation process has been made and a general picture is
emerging.  Star clusters form out of collapsing cloud clumps in
molecular clouds \citep{ti04}.  These collapses are triggered by 
turbulent fragmentation of clouds and their clumps \citep{ma04}.  
The clumps are observed to be aligned in filamentary structures that 
can be reproduced by supersonic turbulent simulations and they contain 
many cores \citep{bu00,kl01}.  Each core forms a single star or a 
binary.  The mass function found for these cores is essentially the 
same as the initial mass function of the stars \citep{jo00,go08}.  The 
cores are themselves clumped in small sub-clumps containing a few to a 
few tens of stars \citep{bo03}.  While the star formation efficiency 
(SFE; i.e.\ the fraction of gas which ends up in the star(s)) in these 
cores is high, the overall SFE measured over the whole molecular cloud 
is very low, of the order of a few per cent \citep{cl04}, and $\leq
40$~per cent in cluster-forming clumps \citep[i.e.\ embedded star
clusters,][]{la03}.

\begin{table*}
  \centering
  \caption{Table of the initial parameters of our simulations. The 
    first line shows the parameters of our standard model.  The 
    columns give the filling factor $\alpha$, the star formation 
    efficiency $\epsilon$ and the number of sub-clumps $N_{0}$, 
    followed by the Plummer radius, the cut-off radius, the total 
    mass, the crossing time and the velocity dispersion of the star 
    forming region.  The next two columns denote the mass in stars and 
    the mass in gas (modelled as analytical background).  Finally we 
    show the Plummer radius, the cut-off radius, the mass, the crossing 
    time and the velocity dispersion of each single clump.}
  \label{tab:param}
  \begin{tabular}{rrrrrrrrrrrrrrr}
    $\alpha$ & $\epsilon$ & $N_{0}$ & $R_{\rm pl}^{\rm sc}$ & $R_{\rm
      cut}^{\rm sc}$ & $M_{\rm pl}^{\rm sc}$ & $T_{\rm cr}^{\rm
      sc}$ & $\sigma_{\rm 3D}^{\rm sc}$ & $M_{\rm star}$ & $M_{\rm
      gas}$ & $R_{\rm pl}$ & $R_{\rm cut}$ & $M_{\rm pl}$ & $T_{\rm
      cr}$ & $\sigma_{\rm 3D}$ \\ 
    & & & [pc] & [pc] & [M$_{\odot}$] & [kyr] & [km/s] & [M$_{\odot}$]
    & [M$_{\odot}$] & [pc] & [pc] & [M$_{\odot}$] & [kyr] & [km/s] \\
    \hline 
    0.05 & 0.32 & 16 & 0.20 & 1.00 & 1000 &  260 & 2.5 & 320 & 680 &
    0.01 & 0.05 & 20.0 & 20 & 1.6 \\ \hline
    0.01 & 0.32 & 16 & 1.00 & 5.00 & 1000 & 2950 & 1.1 & 320 & 680 &
    0.01 & 0.05 & 20.0 & 20 & 1.6 \\
    0.02 & 0.32 & 16 & 0.50 & 2.50 & 1000 & 1043 & 1.6 & 320 & 680 &
    0.01 & 0.05 & 20.0 & 20 & 1.6 \\
    0.10 & 0.32 & 16 & 0.10 & 0.50 & 1000 &   93 & 3.6 & 320 & 680 &
    0.01 & 0.05 & 20.0 & 20 & 1.6 \\ 
    0.20 & 0.32 & 16 & 0.05 & 0.25 & 1000 &   33 & 5.0 & 320 & 680 &
    0.01 & 0.05 & 20.0 & 20 & 1.6 \\
    0.50 & 0.32 & 16 & 0.02 & 0.10 & 1000 &    8 & 8.0 & 320 & 680 &
    0.01 & 0.05 & 20.0 & 20 & 1.6 \\ \hline
    0.05 & 0.32 &  4 & 0.20 & 1.00 & 1000 &  260 & 2.5 & 320 & 680 &
    0.01 & 0.05 & 80.0 & 10 & 3.2 \\
    0.05 & 0.32 &  8 & 0.20 & 1.00 & 1000 &  260 & 2.5 & 320 & 680 &
    0.01 & 0.05 & 40.0 & 15 & 2.3 \\
    0.05 & 0.32 & 32 & 0.20 & 1.00 & 1000 &  260 & 2.5 & 320 & 680 &
    0.01 & 0.05 & 10.0 & 29 & 1.1 \\ \hline
    0.05 & 0.10 & 16 & 0.20 & 1.00 & 1000 &  260 & 2.5 & 100 & 900 &
    0.01 & 0.05 &  6.3 & 37 & 0.9 \\
    0.05 & 0.20 & 16 & 0.20 & 1.00 & 1000 &  260 & 2.5 & 200 & 800 &
    0.01 & 0.05 & 12.5 & 26 & 1.3 \\
    0.05 & 0.25 & 16 & 0.20 & 1.00 & 1000 &  260 & 2.5 & 250 & 750 &
    0.01 & 0.05 & 15.6 & 24 & 1.4 \\
    0.05 & 0.50 & 16 & 0.20 & 1.00 & 1000 &  260 & 2.5 & 500 & 500 &
    0.01 & 0.05 & 31.3 & 17 & 2.0 \\
    0.05 & 0.60 & 16 & 0.20 & 1.00 & 1000 &  260 & 2.5 & 600 & 400 &
    0.01 & 0.05 & 37.5 & 15 & 2.2 \\
    0.05 & 0.70 & 16 & 0.20 & 1.00 & 1000 &  260 & 2.5 & 700 & 300 &
    0.01 & 0.05 & 43.8 & 14 & 2.4 \\
    0.05 & 1.00 & 16 & 0.20 & 1.00 & 1000 &  260 & 2.5 & 1000 & 0 &
    0.01 & 0.05 & 62.5 & 12 & 2.8 \\ \hline 
    0.10 & 0.10 & 16 & 0.10 & 0.50 & 1000 &   93 & 3.6 & 100 & 900 &
    0.01 & 0.05 &  6.3 &  37 & 0.9 \\
    0.10 & 0.20 & 16 & 0.10 & 0.50 & 1000 &   93 & 3.6 & 200 & 800 &
    0.01 & 0.05 & 12.5 &  26 & 1.3 \\
    0.10 & 0.25 & 16 & 0.10 & 0.50 & 1000 &   93 & 3.6 & 250 & 750 &
    0.01 & 0.05 & 15.6 &  24 & 1.4 \\
    0.10 & 0.50 & 16 & 0.10 & 0.50 & 1000 &   93 & 3.6 & 500 & 500 &
    0.01 & 0.05 & 31.3 &  17 & 2.0 \\
    0.10 & 0.70 & 16 & 0.10 & 0.50 & 1000 &   93 & 3.6 & 700 & 300 &
    0.01 & 0.05 & 43.8 &  14 & 2.4 \\
    0.10 & 1.00 & 16 & 0.10 & 0.50 & 1000 &   93 & 3.6 & 1000 & 0 &
    0.01 & 0.05 & 62.5 &  12 & 2.8 \\ \hline 
  \end{tabular}
\end{table*}

The remaining gas does not stay in the newborn star cluster but is
driven outwards by stellar feedback.  In embedded clusters containing
more than a few hundred stars the feedback consists of photo-ionising
radiation, the winds of high-mass stars and finally the on-set of the
first supernova explosions \citep{go97}.  For such clusters the
feedback energy can easily be sufficient to unbind the gas leading to
a gas expulsion phase which is rather short, comparable to the
crossing time of the star cluster \citep{k05}.  Pictures of young
massive star clusters, for example in the central region of the
Antennae \citep[NGC 4038/4039][]{wh99}, reveal that they are already
surrounded by H$\alpha$ bubbles stemming from the gas which was blown
out of the star cluster.  It has been shown that these star clusters
can be as young as $5$--$6$~Myr.  The outflow velocities of the gas
have been measured to be $25$--$30$~km\,s$^{-1}$ \citep{wh99}, which
corresponds to gas-evacuation times of $0.2$~Myr for cluster radii of
$4$~pc.  This is comparable to the crossing time of a
$10^{5}$~M$_{\odot}$ cluster.  As a result of this strong and rapid
mass-loss the star cluster is left out of virial equilibrium.  The
velocities of the stars are too high for the reduced mass of the star
cluster.  Hence, even more mass is lost when stars escape from the
star cluster.  This may finally lead to the complete dissolution of
the cluster \citep[also called 'infant mortality', e.g.][]{gr08,gi08}.
But if the star formation efficiency is as high as 33 per cent or
above, a small bound core remains \citep{go97,a00,bo03a,bo03b} which
may account for the survival of low-mass, Pleiades-type clusters
\citep{kr01}.  \citet{ge01} argue that the SFE has to be larger than
50 per cent to get a bound core but if the stars have almost no
initial velocity dispersion then a SFE of 10 per cent could suffice.
This means that the measurement of the initial velocity distribution
in newborn embedded star clusters will be crucial to find out whether,
and how, star clusters survive.

All these authors imply that a spherical embedded cluster has already
formed before the remaining gas is blown out of the star forming
region.  In this paper we want to investigate if this assumption is
valid.  We start at the stage when the sub-clumps have formed (at the
end-point of hydrodynamical SPH-simulations), but the stars are still
in their formation process and therefore have not started to blow away
the remaining gas.  Once the stars start to shine, their stellar winds
blow away the residual gas.  This gas-removal happens on very short
time-scales, but is longer for more massive gas-cores at a given SFE
\citep{pa08}.  For clusters with masses of less than
$10^{5}$~M$_{\odot}$, this gas-removal time is shorter than the
crossing-time of the embedded star cluster \citep{ba08}.  For massive 
star clusters, the explosion of the first supernova will remove the 
residual gas.  This is in principle a different ansatz than most of 
the numerical studies so far \citep[e.g.][]{go98,sc02,go04,mc07}, which 
start with a (mostly cold) clumpy structure without the gaseous 
background.

In this paper we try to answer the following, related, questions:
Is there sufficient time before the gas is driven out for the
sub-clumps to merge and form a spherical, embedded cluster? If so,
what is the effective SFE of these embedded clusters?  By means of
numerical simulations we investigate the time-scale of this merging
process.

\section{Setup}
\label{sec:setup}

The setup of our simulations is based on a typical outcome of
hydrodynamical simulations of a star forming region inside a giant
molecular cloud \citep{bo03}.  We model the star forming region as a
Plummer sphere with a mass of $M_{\rm sc} = M = 1000$~M$_{\odot}$ and
a Plummer radius $R_{\rm pl}^{\rm sc} = 0.2$~pc.  We assume that in
this cloud stars have formed in small sub-clumps with an overall star
formation efficiency $\epsilon = 0.32$.  This gives us
$320$~M$_{\odot}$ in stars, which we divide into $N_{0} =
16$~sub-clumps of $M_{\rm cl} = 20$~M$_{\odot}$.  Each of these
sub-clumps is modelled as a small Plummer sphere with a Plummer radius
of $R_{\rm pl} = 0.01$~pc and a cut-off radius of $R_{\rm cut} =
0.05$~pc.  These clumps are distributed themselves according to the
Plummer distribution function of the whole gas cloud taking a cut-off
radius of $1$~pc.  Because we do not consider the hydrodynamics of the
gas, the remaining mass in gas is modelled as a time-varying
analytical background Plummer potential. The potential is kept
constant except when stated otherwise.

To put the results of this model (our standard model) into context we
also perform a parameter study, in which we vary $R_{\rm pl}^{\rm sc}$,
$N_{0}$ and $\epsilon$.  For clarity we give the parameters of our
standard model as well as for the comparison runs in
Tab.~\ref{tab:param}. 

The study of the merging behaviour of these sub-clumps suffers heavily
from the low number statistics ($N_{0} = 16$).  For each set of
parameters, we therefore run up to $5$ simulations starting from
different random realisations and calculate the average of the
results.

We use the particle-mesh code {\sc Superbox} \citep{fe00} to perform
the simulations.  This code has the advantage that we can use an
arbitrarily high particle number to model each sub-clump, because the
particles in a collision-less code like {\sc Superbox} represent
phase-space elements rather than single stars.  We therefore model
each clump with 100,000 particles.  Furthermore {\sc Superbox} offers
two levels of high resolution sub-grids which stay focused on each
object (i.e.\ sub-clump) and give us the resolution needed to follow
the merging process in detail and to access the time-scales.

As with all particle-mesh codes, developed to simulate galaxies, the
effects of two-body relaxation in our code are almost completely
suppressed.  However, we note that the internal relaxation times of
our sub-clumps is rather low and of the order of a few kyr ($7300$~yr
in the standard model).  Although this implies that the life-times of
those sub-clumps are very short, they will still be larger than the
time needed to have a first passage through the centre of our star
forming region, i.e.\ the time needed to have interactions with other
clumps and the central merger object.  These interactions are a very
violent process which will either destroy the clumps and distribute
their stars in the central region or reset their internal
clocks.  Since internal relaxation will tend to speed up the
dissolution of a star cluster, by neglecting internal processes we are
therefore prolonging the merging process rather than shortening it.

A study by \citet{se00} showed that the internal structure 
of galaxies does not play a role in the merging time-scales (inside a
galaxy cluster) - only the distribution of galaxies inside the cluster
matters.  This argument should also hold for star clusters in a star
cluster complex and we believe we can also apply it to the problem of
this paper.  The key quantities are the distribution of the relative
velocities of the sub-clumps within the star-forming region and how
much energy they can absorb with each interaction. These quantities
depend only on the total masses and effective radii of the sub-clumps,
and hence are reasonably modelled in our simulations.

Dissolving sub-clumps (due to the fast two-body relaxation) will
distribute their stars preferentially in the centre, i.e.\ in and
around the merger object. This process would also tend to reduce the
time-scale for star cluster formation rather than prolonging it. As
before, our simulations are therefore conservative as they provide an
upper limit to the total amount of time required to form the merger
object.
 
We are aware that there are also a lot of stars which are ejected from
the clumps with high velocities due to close encounters within the
sub-clumps and which are subsequently lost to the forming star
cluster. However, these stars only constitute a small fraction of the
total stellar content of the sub-clumps.

We are therefore confident that the conclusions from our simplified
approach to this problem will also hold for more sophisticated models,
including two-body relaxation, which we plan to carry out in the
future.

To make the comparison with previous work \citep{fe02} easier we try
to adopt the same notation of dimensionless parameters and define as a
measure of the concentration of the sub-clumps in the gas cloud
(filling factor)
\begin{eqnarray}
  \label{eq:alpha}
  \alpha = \frac{R_{\rm pl}^{\rm clump}} {R_{\rm pl}^{\rm sc}}.
\end{eqnarray}
We omit the index $^{\rm clump}$ in the following text.  We
furthermore define a dimensionless time as
\begin{eqnarray}
  \label{eq:time}
  \tau = \frac{t}{T_{\rm cr}^{\rm sc}},
\end{eqnarray}
where $T_{\rm cr}^{\rm sc}$ is the crossing time of the star forming
cloud.

\begin{table*}
  \centering
  \caption{Values for $\delta(\alpha)N_{0}$ for the different choices
    of the parameters used in this study.  The simulations are ordered
    in the same fashion as in Table~\ref{tab:param}.  The first
    three columns show the parameters of the simulations, the fourth
    column the predicted values for $\delta(\alpha)N_{0}$ and the
    fifth and sixth column show the values derived if we fit the
    theoretical formulae directly to the data of our simulations.  The
    errors stated in those columns are the standard deviation of the
    fitting values and do not represent the much larger errors due to
    the low number statistics of the simulations.  The final column 
    gives the time needed to merge half of the sub-clumps $t_{1/2}$, 
    calculated using the fitted value from the slow regime.  The time 
    is given in real time, i.e.\ kyr.}
  \label{tab:delta}
  \begin{tabular}{ccc|c|cc|r}
    $\alpha$ & $\epsilon$ & $N_{0}$ & $\delta(\alpha)N_{0}$ &
    $(\delta N_{0})_{\rm fit}$ & $\eta_{\rm fit}$ & $t_{1/2}$ [kyr]\\ 
    \hline 
    0.05 & 0.32 & 16 & 0.094 & 0.104 $\pm$ 0.010 & 0.076 $\pm$ 0.003
    & 2500 \\ \hline
    0.01 & 0.32 & 16 & 0.036 & 0.067 $\pm$ 0.008 & 0.053 $\pm$ 0.004
    & 44000 \\ 
    0.02 & 0.32 & 16 & 0.054 & 0.047 $\pm$ 0.004 & 0.041 $\pm$ 0.002
    & 22000 \\
    0.10 & 0.32 & 16 & 0.149 & 0.141 $\pm$ 0.011 & 0.094 $\pm$ 0.004
    & 620 \\ 
    0.20 & 0.32 & 16 & 0.241 & 0.380 $\pm$ 0.015 & 0.167 $\pm$ 0.009
    & 87 \\ 
    0.50 & 0.32 & 16 & 0.476 & 2.17 $\pm$ 0.92  & 1.38 $\pm$ 0.59 & 4 
    \\ \hline 
    0.05 & 0.32 &  4 & 0.183 & 0.056 $\pm$ 0.003 & 0.043 $\pm$ 0.002
    & 4600 \\ 
    0.05 & 0.32 &  8 & 0.123 & 0.066 $\pm$ 0.005 & 0.053$\pm$  0.003
    & 5500 \\
    0.05 & 0.32 & 32 & 0.074 & 0.0186 $\pm$ 0.0007 & 0.0180 $\pm$
    0.0007 & 14000 \\ \hline
    0.05 & 0.10 & 16 & 0.094 & 0.0060 $\pm$ 0.0003 & 0.0058 $\pm$
    0.0003 & 43000 \\ 
    0.05 & 0.20 & 16 & 0.094 & 0.0156 $\pm$ 0.0012 & 0.0147 $\pm$
    0.0010 & 17000 \\ 
    0.05 & 0.25 & 16 & 0.094 & 0.0216 $\pm$ 0.0017 & 0.0202 $\pm$
    0.0013 & 12000 \\ 
    0.05 & 0.50 & 16 & 0.094 & 0.1541 $\pm$ 0.0190 & 0.0979 $\pm$
    0.0042 & 1700 \\ 
    0.05 & 0.60 & 16 & 0.094 & 0.2537 $\pm$ 0.0317 & 0.1362 $\pm$
    0.0057 & 1000 \\ 
    0.05 & 0.70 & 16 & 0.094 & 0.394  $\pm$ 0.033  & 0.172  $\pm$
    0.006 & 660 \\ 
    0.05 & 1.00 & 16 & 0.094 & 0.476  $\pm$ 0.027 &  0.282  $\pm$
    0.012 & 550 \\ \hline
    0.10 & 0.10 & 16 & 0.149 & 0.027 $\pm$ 0.003 & 0.025  $\pm$ 0.003
    & 3400 \\ 
    0.10 & 0.20 & 16 & 0.149 & 0.039 $\pm$ 0.001 & 0.029 $\pm$ 0.001
    & 2400 \\ 
    0.10 & 0.25 & 16 & 0.149 & 0.042 $\pm$ 0.008 & 0.039  $\pm$ 0.006
    & 2200 \\
    0.10 & 0.50 & 16 & 0.149 & 0.218 $\pm$ 0.012 & 0.118 $\pm$ 0.002
    & 430 \\ 
    0.10 & 0.70 & 16 & 0.149 & 0.255 $\pm$ 0.021 & 0.123 $\pm$ 0.002
    & 360 \\
    0.10 & 1.00 & 16 & 0.149 & 0.192 $\pm$ 0.019 & 0.106 $\pm$ 0.011
    & 480 \\ \hline
  \end{tabular}
\end{table*}

\section{Theoretical Background}
\label{sec:theo}

A theory of merging star clusters was developed by \citet{fe02}.  They
investigated the merging time-scales and efficiencies of young massive
star clusters forming in confined areas which they called star cluster
complexes.  In their case, the star clusters were merging in their own
potential and not in a background potential of the residual gas.  Here
we want to investigate if this theory also applies in the case of
stellar sub-clumps forming in the potential of the residual gas inside
a giant molecular cloud.

In this section we recapitulate their theory and adapt it to our
problem.  \citet{fe02} established first a velocity criterion for
merging based on the work of \citet{ge83} which reads
\begin{eqnarray}
  \label{eq:velcritold1}
  \frac{1}{2} v_{\rm typ}^{2} & \leq & \frac{G M_{\rm pl}}{R_{\rm pl}}.
\end{eqnarray}
The right-hand side is a value of the order of the internal velocity
dispersion of our sub-clumps, while the left-hand side is of the order
of the clump-velocity dispersion in the star forming region.  Only if
this formula is satisfied should merging between objects be possible.
If we insert now $v_{\rm typ} \approx \sqrt{2} \sigma_{\rm pl}$ and the 
definitions for $\alpha$ and $\epsilon$ we  
derive 
\begin{eqnarray}
  \label{eq:velcritold2}
  \frac{N_{0} \alpha} {\epsilon} & \leq & \frac{32}{3\pi}.
\end{eqnarray}

However, \citet{fe02} noted that in simulations with large values of
$\alpha$ the right-hand side of equation~\ref{eq:velcritold1} has to
be replaced by the unknown values of the merger object.  Thus,
equation~\ref{eq:velcritold1} gives us a threshold for low-mass and
low-$\alpha$ simulations, while in cases with a high filling factor
this criterion fails.

\citet{fe02} identified two regimes of the merging process.  As long
as the mean projected distance between two clusters in the centre of 
the cluster is smaller than the merger radius (i.e.\ high filling 
factors) the merging process is very fast, because the clusters in 
the centre form a central merging object almost immediately and the 
subsequent merging happens preferably with this object which covers a 
fraction $\eta$ of the surface area of the star forming region.  
Therefore the sub-clumps merge following a 'fast' exponential 
decrease in their numbers: 
\begin{eqnarray}
  \frac{{\rm d}N} {{\rm d}\tau} & = & - N \eta \ \Longrightarrow 
  \nonumber \\
  \label{eq:fast}
  N(\tau) & = & N_{0} \exp(-\eta \tau).
\end{eqnarray}
$\eta$ is the fraction of the surface of the star-forming region, 
which is covered by the merger object and leads to a merging of the 
passing sub-clump.  If time is measured in crosing times, i.e.\ the 
time the sub-clumps need to travel once through the star forming 
region, the merger rate can be derived by simple geometrical 
arguments.  We note that $\eta$ replaces the $\epsilon$ from 
\citet{fe02}, which we now reserve for the star formation efficiency.  
\citet{fe02} find that $\eta_{\rm F} = 0.2$ in their models.  This 
value was found to be independent of the choice of $\alpha$ for a wide 
range of values.  

If the concentration of sub-clumps sinks below a threshold, i.e.\
when the mean projected distance between two clusters in the centre is
larger than the merger radius, the 'slow' regime takes over.  The
sub-clumps still merge fast but rather with each other and build up
the final merger object at later stages.  Then the merging follows
\begin{eqnarray}
  \frac{{\rm d}N} {{\rm d}\tau} & = & - N 
  \frac{(N-1) \pi r_{\rm merger}^{2}}{A_{\rm sc}} \ \approx \ 
  \delta(\alpha) N^{2} \ \Longrightarrow \nonumber \\
  \label{eq:slow}
  N(\tau) & = & N_{0} \frac{1} {1+ \delta(\alpha)N_{0}\tau}.
\end{eqnarray}
Here $ r_{\rm merger}$ denotes the merger radius, i.e.\ the maximum 
distance two clumps could have to merge after the interaction, which 
(measured in units of the Plummer radius of the star forming region) 
is a function of $\alpha$.  If we follow the same arguments as 
\citet{fe02}, i.e. we start out from the energy criterion in the 
impulsive approximation developed by \citet{sp58}:
\begin{eqnarray}
  \label{eq:spitzer}
  \Delta E & = & \frac{1}{2} M_{\rm cl} \left( \frac{2G M_{\rm cl}} 
    {r_{\rm merger}^{2}v_{\rm typ}} \right)^{2} \frac{3}{2} r_{\rm c}^{2}
\end{eqnarray}
and insert again $v_{\rm typ} \approx \sqrt{2} \sigma_{\rm pl}$,
$r_{\rm c} = 3/2 R_{\rm pl}$ and take into account that the two
clusters will get gravitationally focussed \citep{bi87} we arrive at
the following expression:
\begin{eqnarray}
  \label{eq:delta}
  \delta(\alpha) N_{0} & = & \left( \frac{2048}{3 \pi^{2}} \right)^{1/2}
  \frac{\alpha} {\gamma^{2}} \left( 1 + \left(  \frac{8192}{27
        \pi^{2}} \right)^{1/4} \frac{1} {\sqrt{N_{0} \alpha}}\right),
  \\
  & = & 8.317 \frac{\alpha} {\gamma^{2}} \left( 1 +
    \frac{2.355} {\sqrt{N_{0} \alpha}} \right), \nonumber
\end{eqnarray}
which gives with $\gamma = 4.0 \pm 0.5$ the values for
$\delta(\alpha)N_{0}$ shown in Tab.~\ref{tab:delta}.  $\gamma$
is the mean of the projected radii of the outermost sub-clump in the
realisations of the initial conditions, in units of $R_{\rm pl}^{\rm
sc}$.  

The change between the very fast and not so fast regime takes place at
a low $\alpha$-value of $< 0.02$.  

Our merger criterion (i.e.\ when do we regard two sub-clumps as merged 
or a sub-clump with the merger object) is set up in the following way: 
if two clumps collide with each other and their centres of density do 
not separate more than a certain radius $r_{\rm max}$ for the rest of 
the simulation, then we regard the two clusters as merged.  We favour 
a distance criterion to an energy criterion because all clusters are
bound to each other to begin with and in later stages we already have
a merger object in the centre, which complicates the process.  This
merger object, which is somewhat larger than the initial clusters is
also the reason to choose a rather large $r_{\rm max}$.  From hundreds
of simulations we deduce that $r_{\rm max} = R_{\rm cut}$, where
$R_{\rm cut}$ is the cut-off radius of our single clumps, is a good
choice for identifying two clumps as merged.  It is small enough that
the two single clumps do not separate from each other anymore and
large enough to account for merger events with the central merger
object as well.

\section{Results}
\label{sec:res}

\begin{figure}
  \centering
  \epsfxsize=9cm \epsfysize=9cm \epsffile{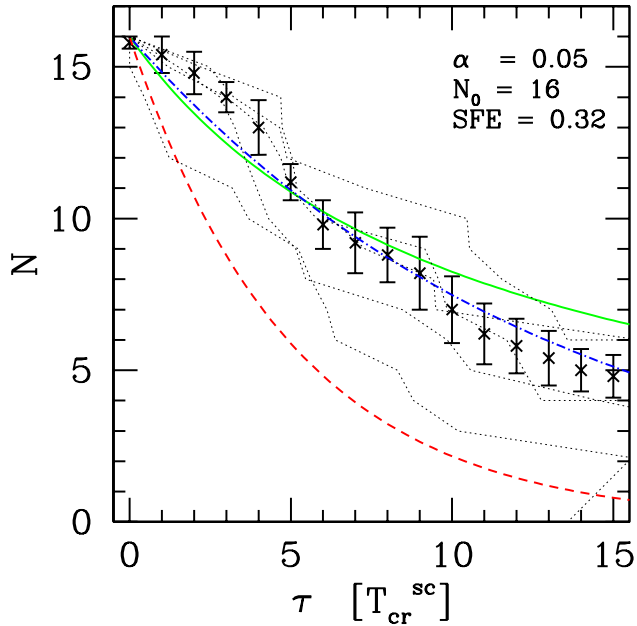}
  \caption{Our standard model.  Data points with error-bars show the
    mean values derived out of $5$ simulations with different random
    seeds.  The results of each of the simulations are shown as dotted
    lines.  Dashed line (red) shows the analytical curve according to
    a 'fast' exponential decrease, solid (green) line shows the 
    analytical curve according to the $\alpha$-dependent 'slow' theory 
    and dot-dashed (blue) line shows the fitted line according to an 
    exponential decrease.}
  \label{fig:mtime}
\end{figure}

\subsection{Standard Model}
\label{sec:standard}

We begin by discussion the results for our standard model.  In
Fig.~\ref{fig:mtime} we show how the number of star clusters decreases
with time.  According to the prediction of \citet{fe02} the number
should decrease exponentially with time ($\alpha > 0.02$).  The data
points can be nicely fit with an exponential decrease with $\eta_{\rm
fit} = 0.076 \pm 0.003$ with a reduced $\chi^{2}$ of $1.00$ (dot-dashed 
(blue) line in Fig.~\ref{fig:mtime}).  This value is lower than the 
prediction of $\eta_{\rm predict} = 0.2$, i.e.\ the merging happens more 
slowly than predicted.  If we use instead the prediction of the 'slow' 
regime we obtain the solid (green) line in Fig.~\ref{fig:mtime}.  This 
line seems to fit the data also.  If we determine the fitted value of
$\delta N_{0}$ for our data, we get $(\delta N_{0})_{\rm fit} = 0.10
\pm 0.01$ with a $\chi^{2}$ of $2.78$.  This value is in very good
agreement with the prediction of the $\alpha$-dependent theory of
$(\delta N_{0})_{\rm predict} = 0.094$.  It seems that the prediction of 
the old theory works well, if we shift the transition value of $\alpha$ 
from the 'slow' to the 'fast' regime to a higher value.  And if we 
check the velocity criterion formula (Eq.~\ref{eq:velcritold2}) we see 
(with our choice of $N_{0}$ and $\epsilon$) that $\alpha \leq 0.07$  
ensures that the sub-clumps merge. 

Going back from our dimensionless time $\tau$ ($T_{\rm cr}^{\rm sc} =
260$~kyr) to the real time $t$, we see that the majority of sub-clumps
merge within $2$--$3$~Myr, i.e.\ faster than any supernova explosion
will occur and blow out the remaining gas from the star-forming
region.  The model shows that by that time one already has a bound, 
nearly spherical merger object in the centre of the star forming 
region. 

We now have to find out why, and where, we have to switch from the
very fast $\alpha$-independent regime to the slower (but still fast)
$\alpha$-dependent theory.  This is also interesting in the light of 
low mass star clusters with $10^3$ to $10^4$ stars, in which the gas 
removal happens even faster through stellar feedback of O~stars.  An 
example of such an object would be the Orion Nebula Cluster which has 
an age of 0.5~to 1.5~Myr and has already blown out most of its natal 
material \citep{kr01}.  We therefore perform a suite of simulations 
with varying $\alpha$, $N_{0}$ and $\epsilon$.
  
\subsection{Parameter Study}
\label{sec:param}

\begin{figure*}
  \centering
  \epsfxsize=8.8cm \epsfysize=8.8cm \epsffile{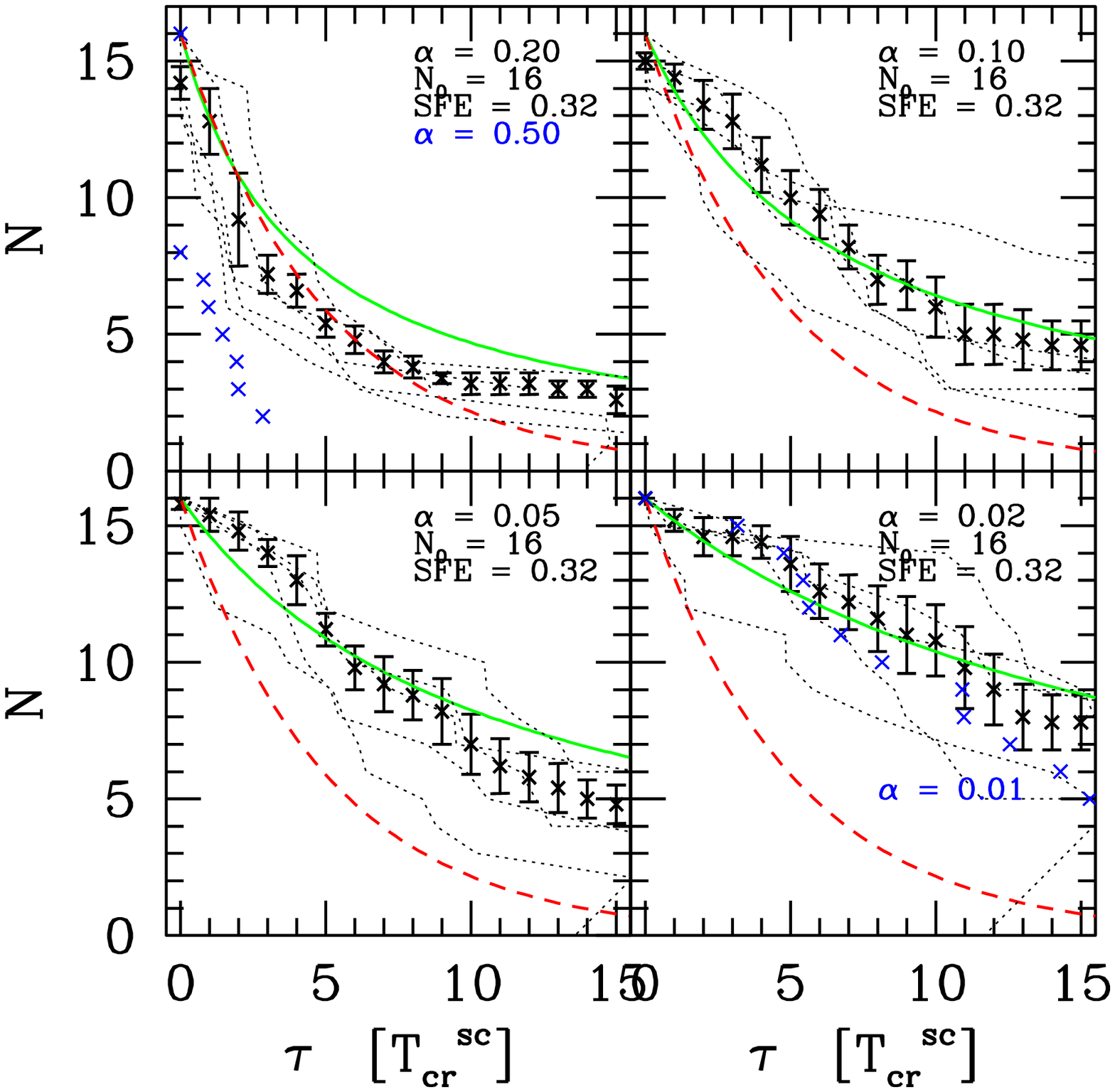}
  \epsfxsize=8.8cm \epsfysize=8.8cm \epsffile{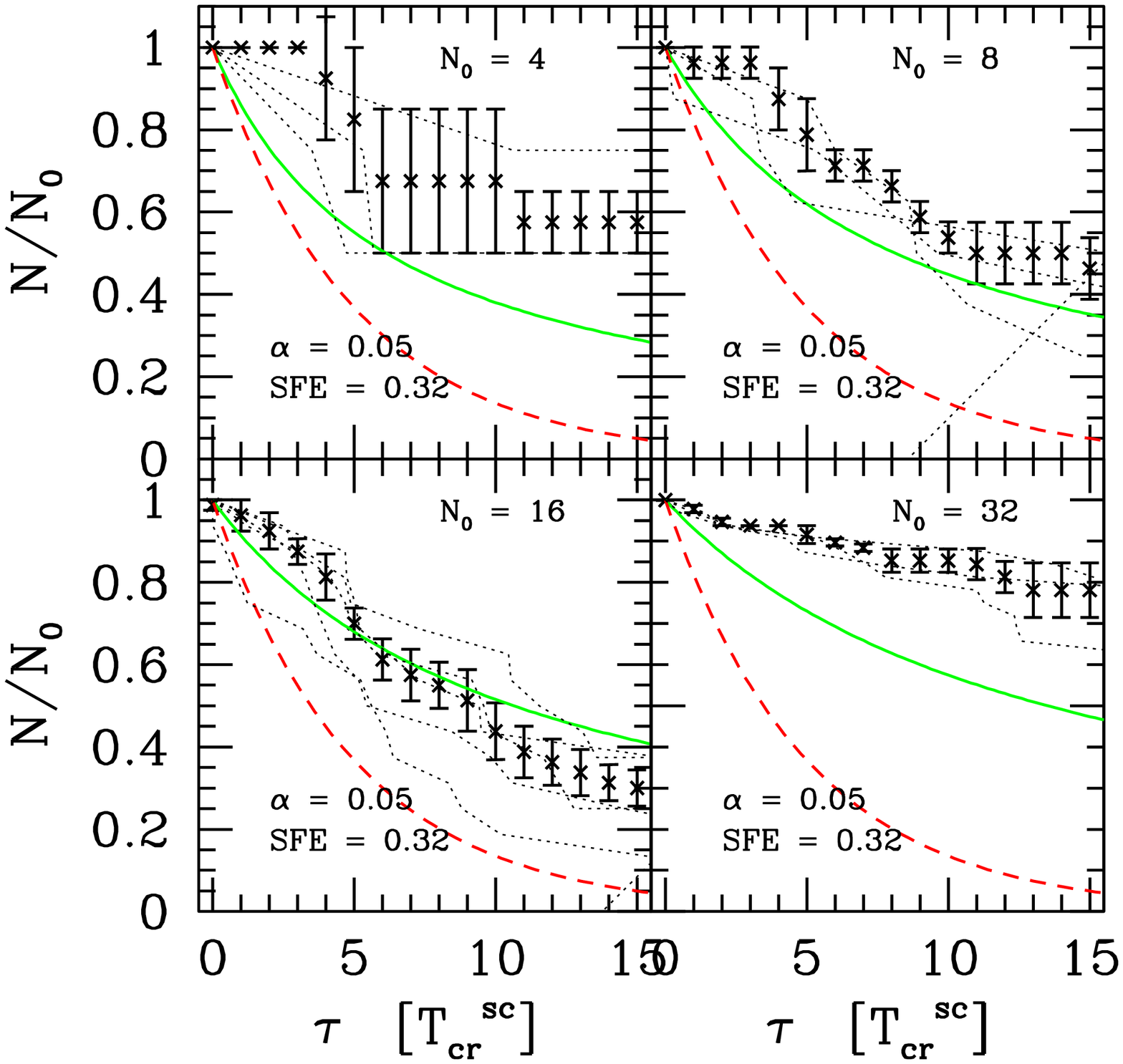}
  \caption{Left: Results of our simulations with varying values
    of $\alpha$.  Solid (green) line shows the predicted decrease
    according to the 'slow' regime, while the dashed (red) curve
    gives the prediction of the 'fast' regime.  The simulations follow
    the 'slow' regime except for very high values of $\alpha \geq
    0.2$.  In the case of $\alpha = 0.5$ (small (blue) crosses) in the 
    top left panel all clumps overlap from the beginning and merge 
    immediately (single simulation).  The light (blue) crosses in lower 
    right panel show a single simulation with $\alpha = 0.01$, which 
    also follow the slow regime.  
    Right: Results of our simulations with varying $N_{0}$.  The 
    results show that the decrease in numbers is fairly independent of 
    $N_{0}$ and that the 'slow' regime (solid (green) lines; dashed (red) 
    lines show the 'fast' regime) fits the data reasonably well 
    (simulations with only $4$ clumps suffer from 'very low number 
    statistics') as long as the number of clumps are below a certain 
    threshold which is reached at $N_{0}=32$.  See main text for 
    explanation.  In both sets of panels dotted lines are the results 
    of the individual simulations.}
  \label{fig:mt2}
\end{figure*}

\begin{figure*}
  \centering
    \epsfxsize=08.8cm \epsfysize=08.8cm \epsffile{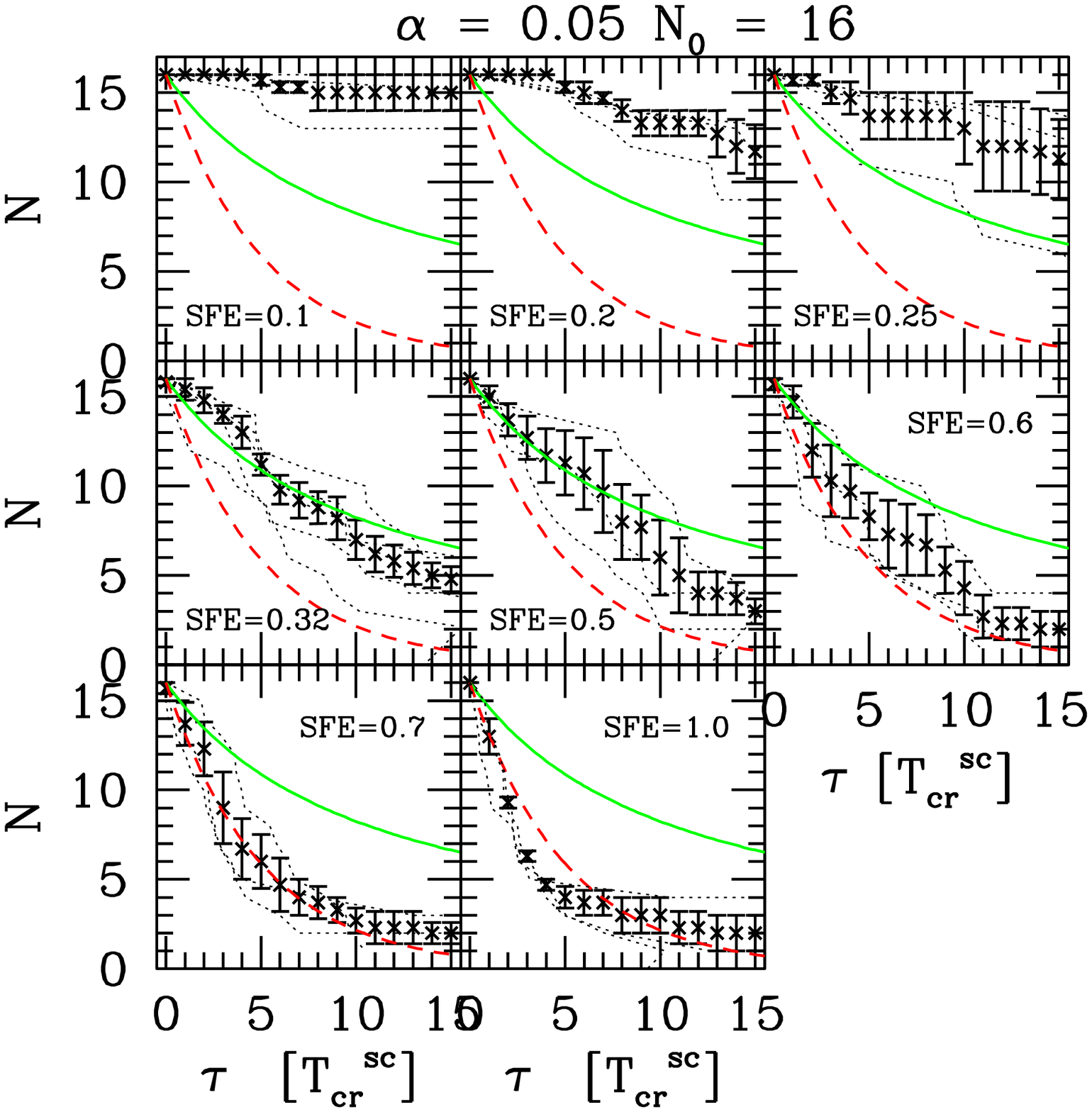}
    \epsfxsize=08.8cm \epsfysize=08.8cm \epsffile{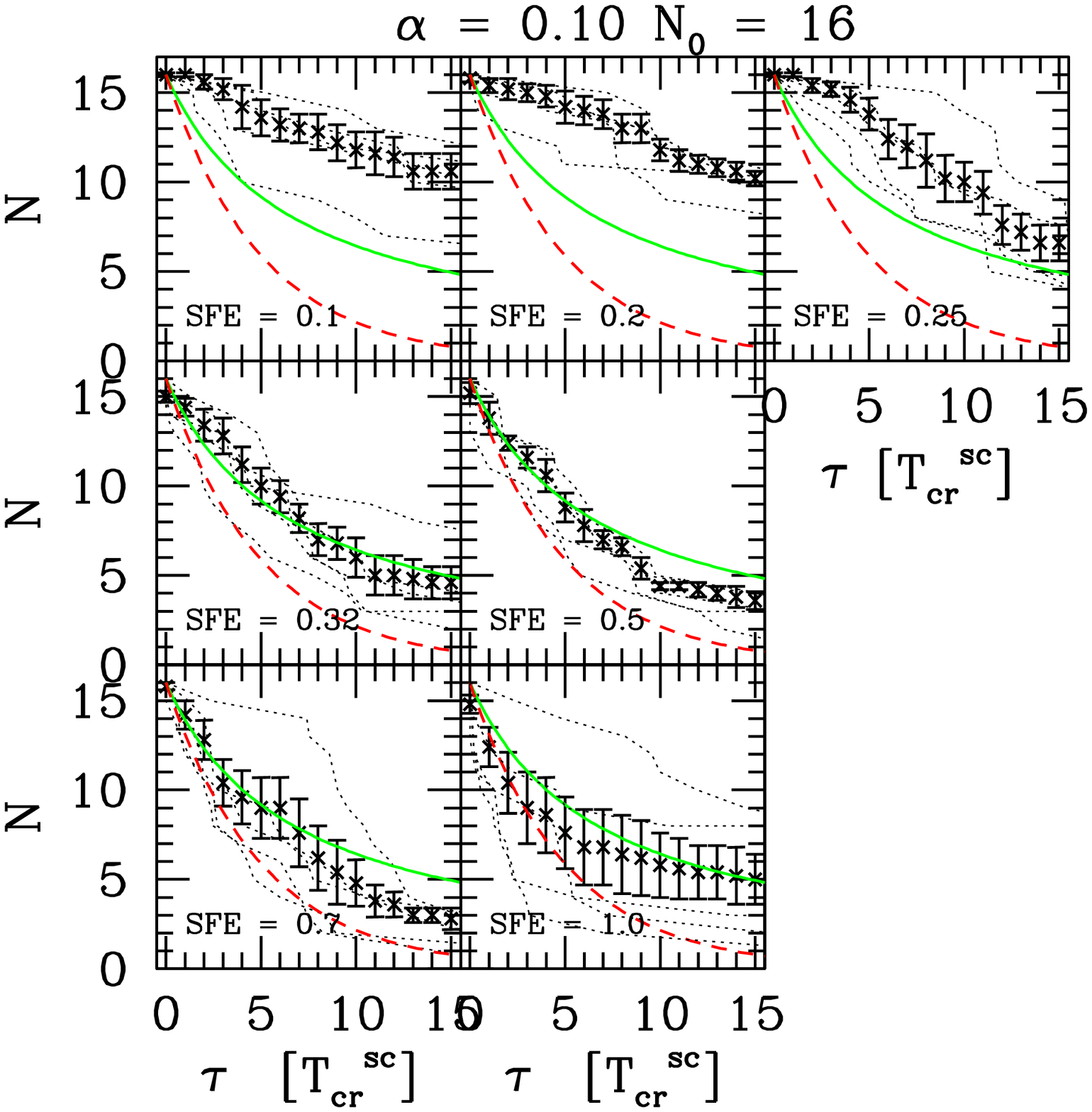}
  \caption{Results of our simulations with different star formation
    efficiencies.  Left: simulations with $\alpha = 0.05$.  Right:
    simulations with $\alpha = 0.1$.  The data points are the mean
    number of clusters after $\tau$ crossing times out of all
    simulations with the same parameters.  Solid (green) line is the
    analytical curve of the 'slow' regime, dashed (red) line is the
    analytical curve for the fast regime.  Dotted lines represent each
    of the single simulations.  One clearly sees that by varying
    $\epsilon$ the results show three different regimes of the merging
    behaviour (see main text for explanation).}
  \label{fig:mt3}
\end{figure*}

In Figs.~\ref{fig:mt2} and~\ref{fig:mt3} we show the results for the
parameter survey.  In the left panels of Fig.~\ref{fig:mt2} we varied
the value of $\alpha$.  It shows that as long as $\alpha \leq 0.2$ the
slow regime fits the data points well.  Only if $\alpha$ becomes very
large $(\approx 0.2)$ does the faster exponential decrease seem to
take over.  Finally, if the filling factor is as large as $\alpha =
0.5$, most of the clumps overlap at the start of the simulation and
the clumps do not separate but stick together and form a cluster from
the start (small (blue) crosses in the top left panel).  This general
behaviour can be understood in terms of our velocity criterion, which
tells us that as long as the filling factor is smaller than $\alpha
\approx 0.07$ merging between the clumps is possible.  But by the time
this threshold is reached, the filling factor is high enough for the
clumps to form a central merger object from the start and the 'fast'
regime takes over.

If we keep $\alpha$ and $\epsilon$ constant and only vary $N_{0}$ the
merging should happen according to the 'slow' regime with only a
slight dependency on $N_{0}$.  The right panels of Fig.~\ref{fig:mt2}
show this behaviour nicely, except for the bottom right panel, when
$N_{0} = 32$.  Here it seems that the clumps do not merge at all.
Again this becomes clear with a look at Eq.~\ref{eq:velcritold2}.
Inserting the constant values of $\epsilon$ and $\alpha$ it results
that $N_{0} \leq 22$ to allow the clumps to merge.  This means by
keeping all parameters constant and dividing the mass in stars into
more and more clumps we reach the point where we leave the regime that
two interacting clumps can absorb enough energy from the orbital
motion that they finally merge.  Merging in this regime is well
suppressed and processes like destruction of the clumps due to the
interactions take over.

Finally we keep $\alpha$ and $N_{0}$ constant and vary the star
formation efficiency $\epsilon$.  The results are shown in
Fig.~\ref{fig:mt3} for two values of $\alpha$.  Using the fitted
values for $\delta N_{0}$ from Tab.~\ref{tab:delta} the data suggest
a strong dependency of $\delta N_{0}$  on $\epsilon$: $\delta N_{0} 
\propto \epsilon^{2}$.  But a closer look reveals that we 
rather have to deal with three different regimes.  The first regime is
governed by a strong background potential and almost no clumps
merging.  Again the mass of a single clump is too low for the clump to
absorb enough energy from the orbital motion into internal energy for
merging to happen, i.e.\ the velocity criterion fails.  If, in the
$\alpha = 0.05$ case (left panels), the star formation efficiency is
higher than $\epsilon > 0.25$, the 'slow' $\alpha$-dependent theory
takes over and fits the data nicely.  This again supports our velocity
criterion which tells us that $\epsilon \geq 0.24$ is required for the 
clumps to be massive enough to allow merging.  If the SFE is higher 
than approximately $\epsilon = 0.5$ then the 'fast' regime takes over, 
i.e.\ the central clumps are able to form a merger object immediately 
and clumps merge with the central object.

In the right panels of Fig.~\ref{fig:mt3} we see the same experiments 
for a higher value of $\alpha = 0.1$.  On a first glance one should 
expect that in this case the merging should happen faster, but the 
velocity criterion tells a different story.  If we increase the filling 
factor by reducing the size of the star forming region (by keeping all 
other quantities constant) we raise the velocity dispersion of the 
clumps in the region and therefore make it more difficult for them to 
merge.  Our simple velocity criterion (Eq.~\ref{eq:velcritold2}) gives 
us a transition value of $\epsilon \geq 0.47$ while in our simulations 
we see that this transition has to be in the region of $\epsilon 
\approx 0.3$.   Also the 'fast' regime is reached only at a higher 
SFE-value of $\epsilon \geq 0.7$.

As result of our parameter study we deduce that even though the
simple geometrical theory of \citet{fe02} does not account for any
background potential it is still applicable, with some limitations.
The velocity criterion of Eq.~\ref{eq:velcritold1} used in
\citet{fe02} gives (translated into the values important in this study
-- Eq.~\ref{eq:velcritold2}) a good hint for which part of the
parameter space the theory, especially the 'slow' regime, is valid.
It still needs to be established when the transition between the
'slow' and the 'fast' regime sets in.  From our parameter study, we 
can conclude that $\alpha$ and $\epsilon$ are anti-proportional to 
each other.  However, the derivation of a general theory to explain 
the precise locations of the transition values of the parameters is 
postponed to a much wider study of the parameter space.

To connect our theoretical study with actual observations we included
in the last column of Tab.~\ref{tab:delta} the real time take for
half of the sub-clumps to merge, which results already in a prominent
merger object, taking the values of Tab.~\ref{tab:param}. This shows
that for our set of parameters for the star forming cloud one can indeed
find configurations that form a merger object even before the stars
manage to expell the gas via stellar winds and very long before the
first supernova is expected. More generally, the way to deduce this
merging time for any given initial conditions is first to calculate
$\delta N_{0}$: the 'half-merger time' in crossing times of the
star forming region is just the inverse of this number. Multiplying
this number with the crossing time in years gives then the final
answer.

\begin{figure*}
  \centering
  \epsfxsize=8.8cm \epsfysize=8.8cm \epsffile{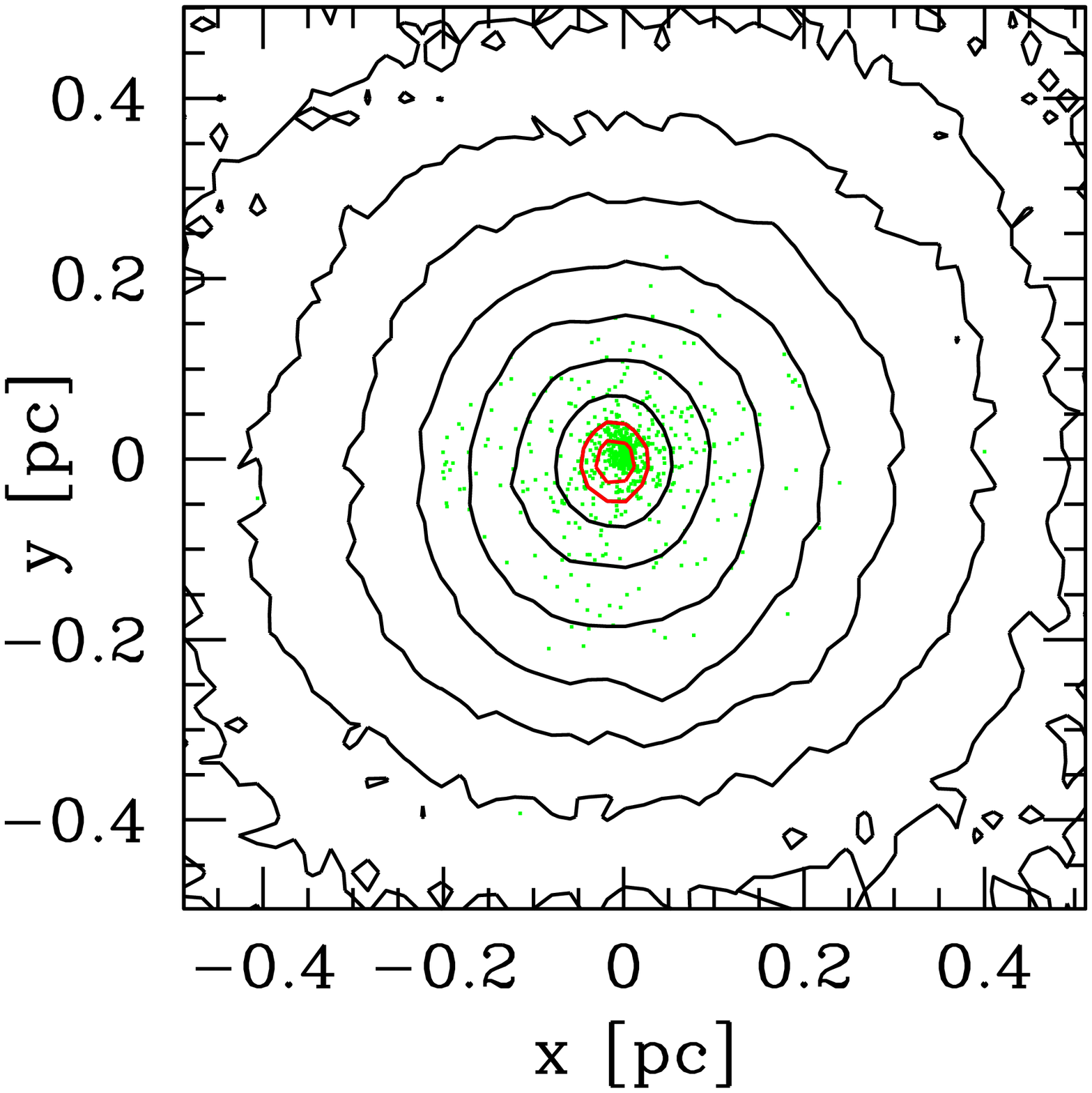}
  \epsfxsize=8.8cm \epsfysize=8.8cm \epsffile{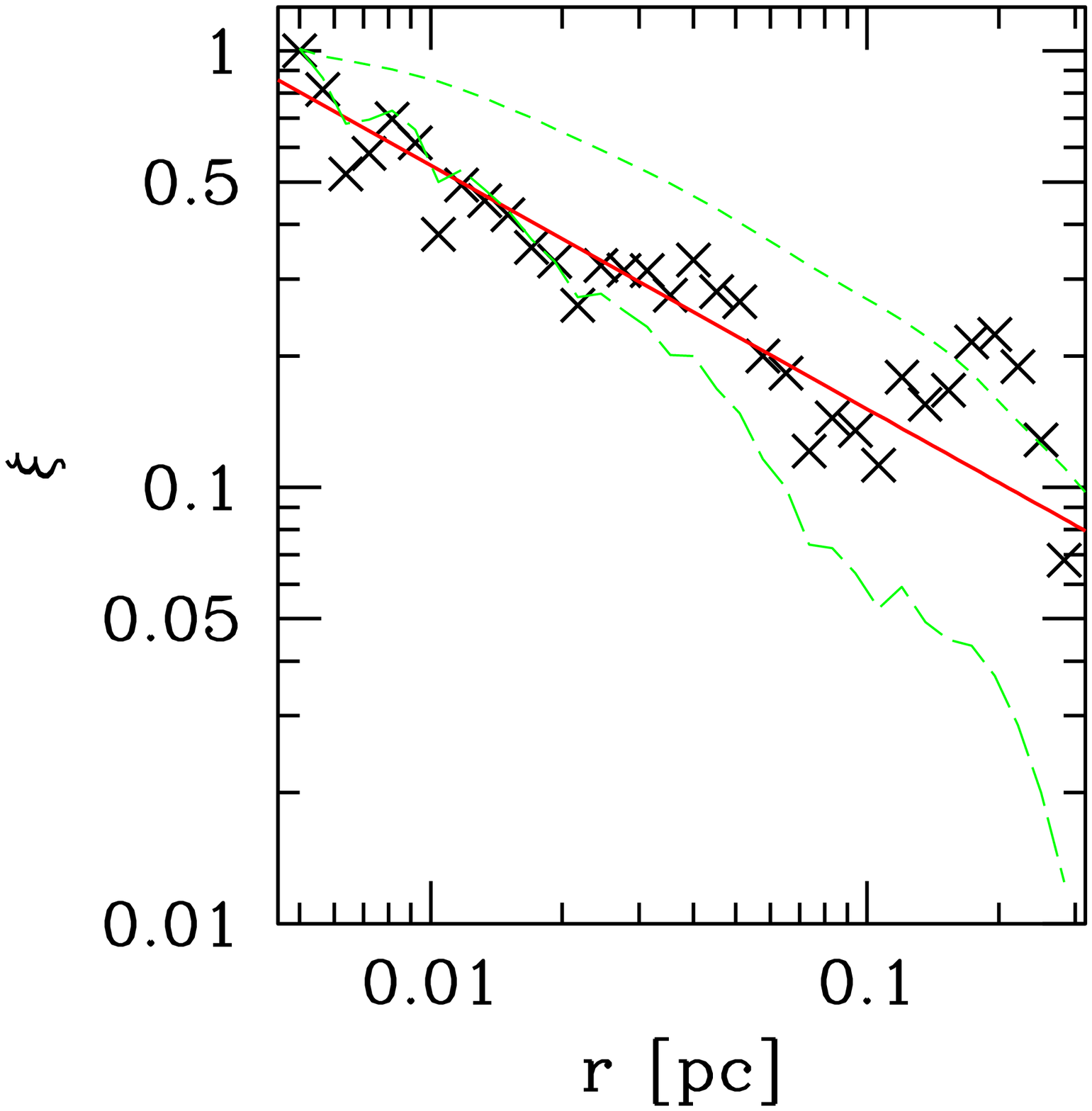} 
  \caption{Left: Contour plot of the final merged cluster of one of our 
    simulations.
    Over-plotted as (green) dots are the positions of the 50
    initially lowest energy ``marked'' stars of each clump.  Right: 
    The ratio $\xi$ (crosses) between the surface-density profile of the 
    marked stars and all stars of the merger object (normalised to $1$) 
    with the solid (red) line showing the power-law fit.  The dashed 
    (green) lines show the two profiles (also normalised to $1$ in the 
    centre) with the short dashed line showing all particles and the 
    long dashed line the lowest (initially) energy particles only.}
  \label{fig:center}
\end{figure*}

As a final remark one should keep the statistical nature of our 
results in mind.  We use only $16$ clumps to represent the distribution 
of clumps in the star forming region.  In such random realisations it 
is very likely that even though we use a large filling factor, i.e.\ a 
high central concentration of clumps, the random numbers give us a 
realisation where most of the clumps are outside the central area and 
therefore act like a distribution with a much lower value of the filling 
factor.  Also the opposite can be true.  Furthermore by repeating the
experiment with the same set of parameters only 3--5 times we get closer 
to a mean representation of our input parameters but we also
introduce extra noise to our study.  In the figure-panels of this 
manuscript we show the behaviour of the single simulations as thin 
dotted lines to illustrate how differently random realisations of the 
same parameters can behave.

\subsection{Mass Segregation}
\label{sec:mass}

We mark the 50 innermost particles ($<1 R_{\rm pl}$) with the lowest 
energy in each sub-clump and locate them 
in the final merger object.  As expected they appear to end up in the
centre of the merger object (see Fig.~\ref{fig:center} top panel).  To
check if this is not merely a visual appearance due to the steep
profile of the merger object, we derived the ratio of the surface
density of the central particles to the surface density of all
particles.  As shown in the middle panel of Fig.~\ref{fig:center} this
ratio ($\xi$; here normalised to unity in the centre) declines as a
shallow power-law with index $n = -0.56 \pm 0.04$.  This shows that
indeed mass-segregated stars in the centre of the sub-clumps end up
again in the centre of the merger object.

This process is a collective effect based on phase-space arguments 
only and is not due to two-body relaxation effects.  Using a 
particle-mesh code for our simulations means that the effects of 
two-body relaxation are almost completely switched off.  The signal 
is smeared out, due to the fact that we only choose particles 
according to their position and not simultaneously if they have a 
velocity which keeps them in the centre, which is the case with 
massive, segregated stars and disruptive effects of close encounters 
of unmerged sub-clumps with the merger-object.  This means that the 
effect that central 
particles of sub-clumps end up in the centre of the merger object is
not a process which requires two-body relaxation but rather a
phase space argument.  Still the results would be more pronounced if
we would have used a direct N-body code, allowing for two-body
relaxation.  This result nicely confirms and even strengthens the
results of \citet{mc07}, who found that mass-segregated clumps form a
mass-segregated star cluster.

\begin{figure}
  \centering
  \epsfxsize=9.0cm \epsfysize=9.0cm \epsffile{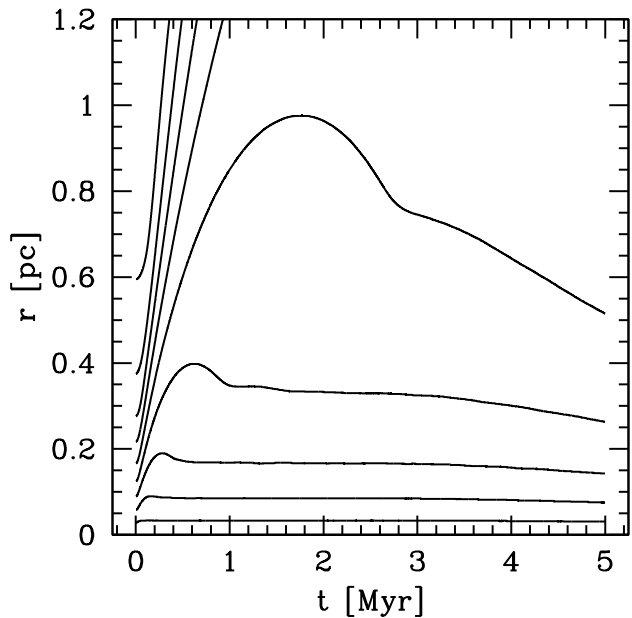}
    \caption{Lagrangian radii of one of our standard 
    simulations after removing the background potential (gas) 
    instantaneously at the start of the simulation.  One
    clearly sees that about 40 to 50 per cent of the cluster remains
    bound.}
  \label{fig:lagrad}
\end{figure}

\subsection{Infant Mortality}
\label{sec:infant}

Infant mortality is one of the major problems in forming star
clusters: How can star clusters survive their gas-expulsion phase
without getting completely destroyed?  Theoretical work predicts that
the star formation efficiency has to be $\epsilon > 0.33$ if the gas
is lost instantaneously to retain a small bound core of a surviving
cluster.  There are several effects which can help to reduce this
transition value and therefore help star clusters to survive their
formation process.

First, if one takes into account internal evolution by simulating the
cluster with an actual initial mass function with heavy stars sinking
very fast to the centre, leading to a compact core, this value can be
reduced slightly \citep[see experiments in][]{bo03a,bo03b}.
Furthermore it takes some time until the gas is removed from the
cluster and the stars adapt to the change in potential.  This helps
especially very small star clusters (without massive stars) and very
massive clusters (long gas-expulsion time).  If star clusters do not
form in isolation but in a star cluster complex the merging of the
star clusters also helps to retain a massive star cluster in the form
of the merger object.

Here we add another mechanism which helps to overcome infant
mortality.  The time-scale of the merging in our standard model is
very fast (even though it is the 'slow' regime).  Within 2--3~Myr we
see a merger object (star cluster) in the centre.  Because the clumps
have formed all over the star forming region and are now together in
the centre, while the gas is decoupled from this merging process, the
'effective' star formation efficiency of the merged star cluster, i.e.\ 
the ratio between stars and total mass inside the merger object (not to 
be confused with the effective SFE as usually found in the literature 
\citet{go08a}), is higher than the overall efficiency of the parent 
cloud.  The star cluster should therefore survive instantaneous 
gas-expulsion better. 

To show this we took the end-point of one of our merger simulations
and removed the background potential instantaneously.  The overall
star formation efficiency of $0.32$ should not allow the star cluster
to survive, especially if modelled with a code which neglects internal
evolution.  But Fig.~\ref{fig:lagrad} shows that on the contrary we
remain with a bound star cluster containing about $40$ to $50$ per
cent of its initial mass.  This can be explained by looking at the
stars-to-gas ratio within the half-mass radius of our merger object.  
Here we see that by keeping the background (gas) potential constant the
'effective' SFE would have been of the order of 44~per cent (compare
with \cite{a00}).

One can now argue that the gravitational pull of the newly formed
star cluster would also collect some of the remaining gas.  But we
already start with a very simplistic realisation of the background
potential which is a strongly centrally concentrated Plummer model,
allowing us to access all quantities analytically, being much more
concentrated than a star forming region would be.  So we indeed
fulfill this criterion from the start.

\section{Conclusions}
\label{sec:conc}

In this study we tried to find out on which time-scales stars forming
in clumps inside a star forming region merge and form an embedded star
cluster.  We have shown that the analytical description developed by
\citet{fe02} for the larger scale of merging star clusters inside a 
cluster complex can be adopted for this problem also.

By introducing a background potential, mimicking the remaining gas,
the basic theory need not be changed at all, although a shift of the
critical values of the parameters which separate particular merging
behaviours is required.  In the old study already a low filling factor
of $\alpha \geq 0.02$ allowed the star clusters to merge very fast,
decreasing their number exponentially.  The presence of a background
potential shifts this limit to rather high $\alpha$ values of about
$0.2$ due to the larger cluster--cluster  velocities.
This can be understood in the sense that the theory itself is
geometrical and velocity independent, because all velocities were
within a certain range which allows for this fast merging.  So at each
crossing time the same fraction of remaining clusters/clumps are able
to merge with the central object.  In the presence of a background 
potential the relative velocities of the clusters/clumps are higher, 
although the clusters have the same ability to absorb energy from the 
fly-bys.  At any given time, this therefore reduces the number of 
clusters with velocities suitable for merging.

Nevertheless, the old theory for the very low $\alpha$ regime
(i.e.\ when the filling factor is so low that merging preferentially
happens between two clusters/clumps at a time), provides a very good
fit to the data.  In this regime the number of clusters decreases
according to $1/1+\tau$.  For our standard model this translates into
a merging time for almost all of the sub-clumps which is much shorter
than the time-scale until the first supernova explodes.  Under these
conditions a virialised star cluster forms well before gas-expulsion,
driven by supernovae, occurs.  For intermediate-mass clusters 
containing between about $10^3$ and $10^4$ stars gas expulsion occurs 
earlier and the requirement that sub-clustering disappears by the time 
the gas is expelled implies stringent constraints on the configuration 
of the sub-clusters.  Nevertheless, such configurations are found in 
our parameter study.  The new, hierarchically formed,
embedded star cluster is then able to survive the gas-expulsion much
more easily than a monolithically formed cluster would have done,
because the 'effective' SFE at its location is much higher than the
overall SFE of the region in which it formed.

Finally this study has confirmed and strengthened the findings of
\citet{mc07} that mass-segregated clumps form mass-segregated
clusters.  
Even though our method does not allow for mass-segregation
at all, we are able to flag the lowest energy  (i.e. the most bound) 
particles in the centre of each clump and
locate them in the final merger object.  We show that these particles
are more centrally concentrated than the other particles.
This shows that this process is governed by simple phase-space
arguments and is not solely due to two-body relaxation.  As we have
shown, it is not an effect of mass-segregation that massive, already
mass-segregated (within their native clump) stars end up in the
central region of the newly formed cluster.
\\

\noindent {\bf Acknowledgements:} MF would like to acknowledge 
financial support from STFC in the UK at the beginning of the project 
and now through FONDECYT, Chile. The authors also would like to thank
C. Clarke for useful comments.  MIW acknowledges the Royal Society
for support.

\label{lastpage}


\begin{thebibliography}{}

\bibitem[\protect\citeauthoryear{Adams} {2000}]{a00} Adams F.C., 2000,
  ApJ, 542, 964

\bibitem[\protect\citeauthoryear{Baumgardt, Kroupa \& Parmentier}
  {2008}]{ba08}
  Baumgardt H., Kroupa P., Parmentier G., 2008, MNRAS, 384, 1231

\bibitem[\protect\citeauthoryear{Binney \& Tremaine}{1987}]{bi87}
  Binney, J., Tremaine, S., 1987, 'Galactic Dynamics', Princeton
  University Press

\bibitem[\protect\citeauthoryear{Boily \& Kroupa}{2003a}]{bo03a}
  Boily C.M., Kroupa P., 2003a, MNRAS, 338, 665

\bibitem[\protect\citeauthoryear{Boily \& Kroupa}{2003b}]{bo03b}
  Boily C.M., Kroupa P., 2003b, MNRAS, 338, 673

\bibitem[\protect\citeauthoryear{Bonnell, Bate \& Vine}{2003}]{bo03}
  Bonnell I.A., Bate M.R., Vine S.G., 2003, MNRAS, 343, 413

\bibitem[\protect\citeauthoryear{Burkert \& Bodenheimer}{2000}]{bu00}
  Burkert A., Bodenheimer P., 2000, ApJ, 543, 822

\bibitem[\protect\citeauthoryear{Clark \& Bonnell}{2004}]{cl04}
  Clark P.C., Bonnell I.A., 2004, MNRAS, 347, L36

\bibitem[\protect\citeauthoryear{de Grijs \& Goodwin}{2008}]{gr08}
  de Grijs R., Goodwin S.P., 2008, MNRAS, 383, 1000

\bibitem[\protect\citeauthoryear{Fellhauer et al.}{2000}]{fe00} 
  Fellhauer M., Kroupa P., Baumgardt H., Bien R., Boily C.M., Spurzem
  R., Wassmer N., 2000, New Ast., 5, 305 

\bibitem[\protect\citeauthoryear{Fellhauer et al.}{2002}]{fe02}
  Fellhauer M., Baumgardt H., Kroupa P., Spurzem R., 2002, CeMDA, 82,
  113 

\bibitem[\protect\citeauthoryear{Gerhard \& Fall}{1983}]{ge83}
  Gerhard O.E., Fall S.M., 1983, MNRAS, 203, 1253

\bibitem[\protect\citeauthoryear{Geyer \& Burkert}{2001}]{ge01}
  Geyer M.P., Burkert A., 2001, MNRAS, 323, 988

\bibitem[\protect\citeauthoryear{Gieles \& Bastian}{2008}]{gi08}
  Gieles M., Bastian N., 2008, A\&A, 482, 165

\bibitem[\protect\citeauthoryear{Goodwin}{1997}]{go97}
  Goodwin S.P., 1997, MNRAS, 284, 785

\bibitem[\protect\citeauthoryear{Goodwin}{1998}]{go98}
  Goodwin S.P., 1998, MNRAS, 294, 47

\bibitem[\protect\citeauthoryear{Goodwin \& Whitworth}{2004}]{go04}
  Goodwin S.P., Whitworth A.P., 2004, A\&A, 413, 929

\bibitem[\protect\citeauthoryear{Goodwin et al.}{2008}]{go08}
  Goodwin S.P., Nutter D., Kroupa P., Ward-Thompson D., Whitworth
  A.P., 2008, A\&A, 477, 823

\bibitem[\protect\citeauthoryear{Goodwin}{2008}]{go08a}
  Goodwin S.P., 2008, replacement for article in "Young massive star 
  clusters - Initial conditions and environments", E. Perez, R. de 
  Grijs, R. M. Gonzalez Delgado, eds., Granada (Spain), September 2007, 
  Springer Dordrecht, 2008arXiv0802.2207G

\bibitem[\protect\citeauthoryear{Johnstone et al.}{2000}]{jo00}
  Johnstone D., Wilson C.D., Moriaty-Schieven G., Joncas G., Smith G.,
  Gregersen E., Fich M., 2000, ApJ, 545, 327

\bibitem[\protect\citeauthoryear{Klessen \& Burkert}{2001}]{kl01}
  Klessen R., Burkert A., 2001, ApJ, 549, 386

\bibitem[\protect\citeauthoryear{Kroupa}{2005}]{k05} Kroupa P., 2005,
  The Three-Dimensional Universe with Gaia, 576, 629
  (astro-ph/0412069)

\bibitem[\protect\citeauthoryear{Kroupa, Aarseth \&
    Hurley}{2001}]{kr01}   
  Kroupa P., Aarseth S., Hurley J., 2001, ApJ, 549, 386

\bibitem[\protect\citeauthoryear{MacLow \& Klessen}{2004}]{ma04}
  MacLow M., Klessen R., 2004, RvMP, 76, 125

\bibitem[\protect\citeauthoryear{Lada \& Lada}{2003}]{la03}
  Lada C.J., Lada E.A., 2003, ARA\&A, 41, 57

\bibitem[\protect\citeauthoryear{McMillan, Vesperini \& Portegies
    Zwart}{2007}]{mc07} 
  McMillan S.L., Vesperini E., Portegies Zwart S.F., 2007, ApJL,
  655, L45

\bibitem[\protect\citeauthoryear{Parmentier et al.}{2008}]{pa08}
  Parmentier G., Goodwin S.P., Kroupa P., Baumgardt H., 2008, ApJ, 
  678, 347

\bibitem[\protect\citeauthoryear{Scally \& Clarke}{2002}]{sc02}
  Scally A., Clarke C., 2002, MNRAS, 334, 156

\bibitem[\protect\citeauthoryear{Sensui, Funato \& Makino}{2000}]{se00}
  Sensui T., Funato Y., Makino J., 2000, PASJ, 51, 1

\bibitem[\protect\citeauthoryear{Spitzer}{1958}]{sp58}
  Spitzer Jr.\ L., 1958, ApJ, 127, 17

\bibitem[\protect\citeauthoryear{Tilley \& Pudritz}{2004}]{ti04}
  Tilley D.A., Pudritz R.E., 2004, MNRAS, 353, 769

\bibitem[\protect\citeauthoryear{Whitmore et al.}{1999}]{wh99}
  Whitmore B.C., Zhang Q., Leitherer C., Fall S.M., 1999, AJ, 118,1551

\end{thebibliography}
\end{document}